\documentclass{elsart}

\usepackage{epsfig}

\usepackage{amssymb}

\begin{document}

\begin{frontmatter}



\title{Gamma-rays from Muon Capture in $^{14}$N}


\author{T.J. Stocki\thanksref{tr1}},
\author{D.F. Measday},
\author{E. Gete\thanksref{tr2}},
\author{M.A. Saliba\thanksref{tr3}},
\author{B.A. Moftah\thanksref{tr4}},

\address{Department of Physics and Astronomy \\
         University of British Columbia \\
         Vancouver, B.C., Canada V6T 1Z1}

\author{T.P. Gorringe}
\address{Department of Physics and Astronomy \\
         University of Kentucky \\
         Lexington, KY, U.S.A. 40506-0055}

\thanks[tr1]{Corresponding author now at: Communications Research Centre \\
                     3701 Carling Avenue, Box 11490, Station H \\
                     Ottawa, Ontario, Canada K2H 8S2 \\
                     phone: (613) 990-5810\\
                     fax: (613) 993-7139 \\
                     email: trevor.stocki@crc.ca }
\thanks[tr2]{now at:   London Regional Cancer Centre \\
                       London,Ontario, Canada N6A 4L6 }

\thanks[tr3]{now at: Department of Manufacturing Engineering \\
                     University of Malta, Msida  MSD 06, Malta }

\thanks[tr4]{now at: Montreal General Hospital \\
                     McGill University Health Centre \\
                     Montreal, Qu\'{e}bec, Canada H3G 1A4 \\}

\begin{abstract}
   Many new $\gamma$-rays have been observed, following muon capture on
$^{14}$N.  One had been reported before, and the low yield is confirmed,
indicating that the nuclear structure of $^{14}$N is still not understood.
Gamma-rays from $^{13}$C resulting from the reaction
$^{14}$N($\mu^{-}$,$\nu$n)$^{13}$C compare favourably with states observed
in the reaction $^{14}$N($\gamma$,p)$^{13}$C.  More precise energies are also
given for the 7017 and 6730 keV $\gamma$-rays in $^{14}$C.
\end{abstract}

\begin{keyword}
muon capture \sep $^{14}$N \sep gamma-rays
\PACS 25.30.-c \sep 23.40.Hc \sep 27.20.+n 
\end{keyword}
\end{frontmatter}

\section{Introduction}
\label{}
Muon capture had been shown to be an excellent reaction for studying isovector
magnetic transitions in light elements.  The weak interaction aspects of the 
reaction are well understood, at the level of a few percent; this is far better
than the nuclear structure uncertainties, which can be a factor of 2, or even
greater.  A recent survey was carried out by Gorringe {\it et al.}\cite{r1,r2}
for
the targets of $^{23}$Na, $^{24}$Mg, $^{28}$Si, $^{31}$P, and $^{32}$S.  They
showed that the muon capture transition rates are similar to those observed
in the (n,p) reaction, but cannot be reproduced very well by calculations
based on the OXBASH code \cite{r3} for nuclear structure in the 2s-1d shell.
Brown
and Wildenthal \cite{r4} obtained the universal SD interaction, used in this
code, by fitting 440 energy levels in the 2s-1d shell. \\

For 1p shell nuclei the situation is better on the whole.  Muon capture on 
$^{12}$C has been reproduced by the recent calculations of Hayes and Towner
\cite{r5} and Volpe {\it et al.}\cite{r6a}, and $^{16}$O seems to be reasonably
understood too, as indicated by the work of Haxton and Johnson\cite{r6} and
Warburton  {\it et al.}\cite{r6b}.  However these are recent
calculations using quite complex descriptions of the A=12 and A=16 nuclei.  The
case of $^{14}$N remains one of the more puzzling examples amongst light
nuclei.  The experiment of Giffon {\it et al.}\cite{r7} is the most recent
experiment
of muon capture on $^{14}$N, and they observed only one $\gamma$-ray, viz the
one at 7017 keV, which is the ground state transition from the 7019 keV level
(the difference in energy is because the recoil energy is quite significant
for such a high energy gamma and a light nucleus).  Now Giffon
{\it et al.}\cite{r7}
obtained a capture rate to this level of 4640 $\pm$ 700 s$^{-1}$ which is 
equivalent to a yield of (7.0 $\pm$ 1.2) \% per muon capture.  Earlier 
measurements of this capture rate were slightly higher but of lower accuracy,
and will be discussed later.  We take the total muon capture rate to be
(66 $\pm$ 5) $\times$ 10$^{3}$ s$^{-1}$, taking an average of three
inconsistent measurements \cite{r8}.\\

There are three calculations of the capture rate to the 7019 keV level;
Mukhopadhyay \cite{r9}
obtained a value of 20,000 s$^{-1}$ (i.e. a yield of 30
\% !!).  He used the Cohen-Kurath model for 1 p shell nuclei, and harmonic
oscillator radial wave functions.  Mukhopadhyay realized that the calculated
rate was unrealistic and so estimated that the (2s-1d)$^{2}$ excitations in
the 7019 keV level could reduce the rate to $\approx$ 9700 s$^{-1}$, which is
better but still somewhat high.  A later calcuation by Desgrolard
{\it et al.} \cite{r10} confirmed that in the Cohen-Kurath model, the capture
rate is $\approx$ 23,000 s$^{-1}$, and the addition of exchange currents
reduces the
rate by only 20 \% which is far from sufficient \cite{r11}.  Thus it is clear
that for
$^{14}$N, these calculations are in serious disagreement with experiment even
though there is no major problem with other nuclei (apart from $^{11}$B which
is also an ongoing problem \cite{r12}).  These calculations were for the
allowed transitions, i.e. the 1$^{+}$ transitions which, from the 1$^{+}$
ground state of $^{14}$N, excite the $^{14}$C ground state (0$^{+}$), the 6589
keV level (0$^{+}$) and the 7019 keV level (2$^{+}$), under discussion.  Two
other unbound levels are also fed, a 2$^{+}$ at 8.3 MeV and a 1$^{+}$ at 11.3
MeV with a calculated rate of $\approx$  1.3 $\times$ 10$^{3}$ (yield of
$\approx$ 2 \%). \\

Many other reactions excite these same levels; for example (d,$^{2}$He),
($\pi^{-}$,$\gamma$), and (n,p).  The best comparison is probably the
(d,$^{2}$He) reaction, but unfortunately no data exists.  The
($\pi^{-}$,$\gamma$) reaction has been studied with an energy resolution of
720 keV \cite{r13} and clear structure is observed at 7.0, 8.3, 10.7, 15.4,
and 20 MeV
excitation energy.  Only the first peak is bound and thus all the possible
transitions to the 6 excited bound states are not resolved.  The
($\pi^{-}$,$\gamma$) reaction has been reviewed in general by Gmitro
{\it et al.}\cite{r14} and
the reaction $^{14}$N($\pi^{-}$,$\gamma$) in particular has been analyzed
by Kissener
{\it et al.} \cite{r15,r16}.  Now the ($\pi^{-}$,$\gamma$) reaction proceeds
mainly from the
2p atomic state and only a small fraction from the 1s state, depending on the
mass; for $^{14}$N R$_{p}$:R$_{s}$ = 5:1. Because the 2p
state adds a 1$^{-}$ to the initial state J$^{\pi}$ conditions, 2$^{+}$ and
3$^{+}$ transitions are more prominent.  Thus a direct comparison between the
($\pi^{-}$,$\gamma$) and ($\mu^{-}$,$\nu$) reactions is hazardous, but if a 
calculation has addressed both reactions, one can have more confidence in that
approach.  Similarly the (n,p) reaction can be used for another valuable 
comparison.  The reaction $^{14}$N(n,p)$^{14}$C has been studied at 59 MeV
by Needham {\it et al.},\cite{r17} and at 280 MeV by a group at TRIUMF 
\cite{r17a}, both with an energy resolution of about 1 MeV.  Together these
experiments give an excellent and consistent picture of the situation.  The 
TRIUMF data at 0$^{\circ}$ show levels at 7.0 MeV (2$^+$, 46\%); 8.3 MeV
 (2$^+$,36\%); 10.4 MeV (2$^+$,8\%); and 11.3 MeV (1$^+$,10\%).  The J$^{\pi}$
of each 
level is given, as is the contribution to the 0$^{\circ}$ strength
(roughly B(GT)). Note that the group also has complementary and consistent
data on the (p,n)
and (p,p$^{\prime}$) reactions.  The Needham {\it et al.} results are
compatible, with figures illustrating data at 16$^{\circ}$, 18$^{\circ}$, and
48$^{\circ}$, which show levels excited at 7.0, 8.3, 11.3, 15.4 MeV and a 
double peaked giant resonance at about 20.4 MeV.  The angles of 16$^{\circ}$
and 18$^{\circ}$ correspond best to a momentum transfer of the ($\mu^{-}$,
$\nu$) reaction, although 59 MeV is a little low in bombarding energy.  The
main conclusion from all these reactions is that the 1$^+$ transition
strength is much more distributed than that calculated in the Cohen-Kurath
model. \\

We thus come to the final calculation of the $^{14}$N($\mu^{-}$,$\nu$)$^{14}$C
reaction, that by Kissener {\it et al.}\cite{r18}.  They were analyzing both
the
($\mu^{-}$,$\nu$) and the ($\pi^{-}$,$\gamma$) reactions, and hence were more
aware of the pitfalls.  In particular they knew that (e,e$^{\prime}$)
scattering had shown that
the 1$^{+}$ strength to 2$^{+}$ levels was split about equally between the
bound level at 7019
keV, and an unbound one at 8.3 MeV.  In the Cohen-Kurath model, the bound
level takes almost all the strength.  However Kissener {\it et al.} split the
strength, apparently phenomenologically, and not by introducing 2p-2h
excitations in their calculation.  This calculation resulted in a yield of
11\% to each of
these levels, and explains at least one factor of 2.  The other important
feature of their calculation is that they address all the possible transitions,
including high energy excitations in $^{14}$C which decay via neutron emission
to levels in $^{13}$C and $^{12}$C.  Although the results of the calculations
do not correspond in every detail to the observations in the present
experiment, at least a good overall picture is given.  Thus for example
Kissener {\it et al.} show that the $^{14}$C ground state is only weakly fed
in direct transitions by the ($\mu^{-}$,$\nu$) reaction, and this is related to
the slow beta decay rate of $^{14}$C.  Kissener {\it et al.} also find that
the sum
of transitions to bound levels of $^{14}$C is 15\%; for higher energy 
excitations in $^{14}$C, they estimate that the yield is 42\% to bound levels
of $^{13}$C, and 40\% to bound levels of $^{12}$C.  The rest is made up of 
charged particle emission ($\approx$ 3\%), which is probably a low value
($\approx$ 15\% is more common).  Note that Kissener {\it et al.} calculated
the
total muon capture rate to be 109 $\times$ 10$^{3}$ s$^{-1}$ compared to the
experimental value of (66 $\pm$ 5) $\times$ 10$^{3}$ s$^{-1}$, but by 
presenting their results as yields per capture, they circumvent that problem.
\\

For the reaction $^{14}$N($\mu^{-}$,$\nu$n)$^{13}$C, one can envisage two
reaction modes.  One is a direct knock-out term, similar to the pole-term seen
in the ($\pi^{-}$,$\gamma$) reaction.  The levels observed in $^{13}$C would
thus correspond to levels excited in reactions such as (d,$^{3}$He),(p,2p)
or (e,e$^{\prime}$p).  The spectroscopic factors for these reactions are very
similar.
  An alternative description of muon capture would be in terms of an excitation
of a spin-dipole giant resonance (mainly 2$^{-}$ and some 1$^{-}$), followed
by a de-excitation similar to the ($\gamma$,p) reaction.  Of course the 
($\gamma$,p) reaction goes via the E1 giant dipole resonance, which is at a
similar but not identical energy.  The comparison between the knock-out and
excitation models has been discussed in some detail in a recent review of muon
capture \cite{r19}.  There is not a lot of difference between the observable
effects, but as long ago as the muon capture work of Miller
{\it et al.} \cite{r20} in 1972 on $^{28}$Si, it was shown that the
($\gamma$,p) reaction
is a better analogue.  Of course there will be some knock-out component, 
especially when a high energy neutron is produced in muon capture, but this 
represents only 10 \% to 20 \% of neutron production.  The data for the 
reaction $^{14}$N($\gamma$,p) are limited, but we shall make a fruitful
comparison. \\

One unresolved problem in muonic nitrogen is the hyperfine effect.  For nuclei
with spin, the 1s state in a muonic atom is split into two hyperfine levels,
which can be a few eV to $\approx$ 1 keV apart in energy.  The M1 transition
is too slow to be important, but Telegdi and Winston and co-workers at Chicago
\cite{r21,r22} showed that Auger emission can speed up the transition rate to
time scales of $\approx$ $\mu$s in light elements, and $\approx$ ns in heavy
elements.  The best example is $^{19}$F which was studied by the Chicago group
and has been confirmed several times since \cite{r23}.  The case of $^{14}$N
is quite perplexing because a depolarization was observed in $\mu^{-}$SR by
Ishida {\it et al.}\cite{r24}.  Although some of the effect was attributed to 
external electromagnetic fields, the residual was ascribed to a hyperfine
transition rate of (0.076 $\pm$ 0.033) $\mu$s$^{-1}$, i.e. $\tau$ $\approx$
13 $\mu$s.  Now Winston\cite{r22}
calculated the muonic hyperfine rate, via a simple atomic model of electron
ejection, and his model fits every known example, especially $^{19}$F.  For
$^{14}$N the prediction is that there should be no effect, because the
separation of the hyperfine energy levels is 7.4 eV, whereas in the carbon 
atom, the least bound electron is bound by 11.3 eV, i.e. the ionization
potential.  (the $\mu^{-}$ is well within the inner electrons, so the muonic
$^{14}$N system appears to be a $^{14}$C nucleus to the atomic electrons).
If the pseudo $^{14}$C atom has formed a ``C''N bond, the ionization potential
is 14.3 eV, which makes Auger emission even less likely.  Thus it is possible
that Ishida {\it et al.}\cite{r24} observed other depolarization mechanisms.
We note that Wiaux \cite{r12}, using a muon capture technique for $^{11}$B,
observed a slower hyperfine transition rate than that observed in a $\mu^{-}$SR
experiment.  We have searched for hyperfine effects in $^{14}$N, but because
the supposed time constant is long, it is difficult
to detect, and our observations are inconclusive.  Thus we shall assume that
our capture rates are for a statistical mixture of the two hyperfine states.  
However even if there were a slow hyperfine transition, the effect on our
results would be smaller than the errors. \\

Thus muon capture in $^{14}$N exhibits many interesting effects.  Our
experiment was proposed in order to obtain some more relevant observations.
One great advantage of $^{14}$N is that there are only a few bound levels in
$^{14}$C, $^{13}$C, and $^{12}$C.  Thus we are able to make a complete analysis
of all possible transitions for these nuclei.

\section{Experimental Method}
\label{}
The experimental technique is well established, and has been described in
several previous publications \cite{r1,r2}.  Details can be found in the thesis
of Stocki\cite{r23}.  We shall limit ourselves to the more important features.
\\

The experiment was performed on the M9B channel at the TRIUMF cyclotron.  The
beam line includes a 6 m, 1.2 T superconducting solenoid in which 90 MeV/c
pions
can decay.  The resulting backward muons are then selected by a bending magnet
and pass through a collimator into the experimental area.  The collimator was
made out of lead bricks, but faced with 13 mm of polyethylene to minimize
background neutrons and $\gamma$-rays.  The beam rate was about 2 $\times$ 
10$^{5}$ s$^{-1}$, with a pion contamination of $<$ 0.2 \% and an electron
contamination of $\approx$ 20 \%.  Three plastic scintillators defined a muon
stop, the defining counter being 51 mm in diameter. \\

The target was liquid nitrogen contained in a styrofoam box, internal
dimensions 215 mm (beam direction) $\times$ 195 mm $\times$ 550 mm.  The length
in the beam direction was more than sufficient to stop the muons, so the veto
counter behind the target was effectively redundant.  A mu-metal shield was
immersed in the liquid nitrogen to reduce the ambient magnetic field from 1.5
gauss to 0.1 gauss.  This minimized spin precession effects. \\

To detect the $\gamma$-rays from muon capture, two high purity germanium 
detectors (HPGe) were placed opposite each other at 90$^{\circ}$ to the beam
direction.  Both detectors were n-type detectors to minimize radiation damage.
The largest was Ge1, a 44\% detector, with an in-beam energy resolution of
2.53 keV at 1.3 MeV and a timing resolution of 6 ns.  The other, Ge2, was a
21\% detector, with an in-beam energy resolution of 2.19 keV at 1.3 MeV, and a
timing resolution of 7 ns.  Their front windows were 27.5 cm and 33.1 cm away
from the centre of the target, respectively.  In front of each was a plastic
scintillator to tag electron events.  Each HPGe was surrounded by a Compton
suppressor, composed of a segmented NaI(Tl) annulus.  In this experiment the
suppressors were implemented in software mode, so single and double escape
peaks were in fact observed, so we were able to take advantage of that
feature. \\

The electronics consisted of spectroscopic amplifiers and timing filter
amplifiers, followed by constant fraction discriminators (CFD).  An event was
defined by a pulse in a germanium detector, and then tags from the electron
counters, and pulse height information from the Compton Suppressors were also
recorded.  Events from a delayed muon stop were then sent to a router for 10.8
$\mu$s.  This device can accept up to 4 pulses and route them to four distinct
TDCs.  Thus one can identify events with only one candidate muon, from those
with two, three, or even four muon stops in the previous 10.8 $\mu$s.  
Typically 15\% of events have two muon stops, which made selection of events
problematic, for determination of lifetimes.  Two methods were used; one
selecting events with only one muon stop, the other accepting all events.  
Accepting the first muon, whether or not it was followed by another muon,
created a clearly distorted time spectrum. \\

The energy calibration of the detectors was carried out by a variety of 
techniques.  Offline $^{152}$Eu and RdTh sources were useful.  The
$\gamma$-ray energies were known to a better than 10 eV which is far more 
accurate than is needed.  Of course the beam on conditions are somewhat
different, and calibrations at a much higher energy were required.  The
spectra were divided into 3 regions (low, medium, and high energy) and it was
found that, in each region, linear fits for the energy calibration were quite
adequate.  The beam-on calibrations that were used are listed in
 Table~\ref{tab1}.
Note that the (n,$\gamma$) lines are room background and not due to the liquid
nitrogen target, but they are narrow lines and therefore are convenient for
calibration purposes, especially as the energies are known to 0.1 keV in a 
region with few calibration lines.  Also noticeable is the $\gamma$-ray from
$^{16}$N $\beta$-decay, coming from the irradiated cooling water in the nearby
quadrupoles.  It has an energy of 6,129,140(30) eV\cite{r33} whereas several
compilations seem to use the erroneous excitation energy of E$_{x}$ =
6,129,893(40) eV and a $\gamma$-ray energy of 6,128,630(40) eV \cite{r34,r35}.
We note that we have used the single and double escape peaks with energy 
differences of 511.00 and 1022.00 keV respectively.  This is unreliable and
the actual difference can be 0.1 to 0.3 keV less, depending on the detector
\cite{r32,r33,r34,r35,r36}.  Because our lines are Doppler broadened, such
small calibration errors were not studied.  Higher precision work would need
to take this effect into account. \\

\begin{figure}
\begin{centering}
\epsfig{figure=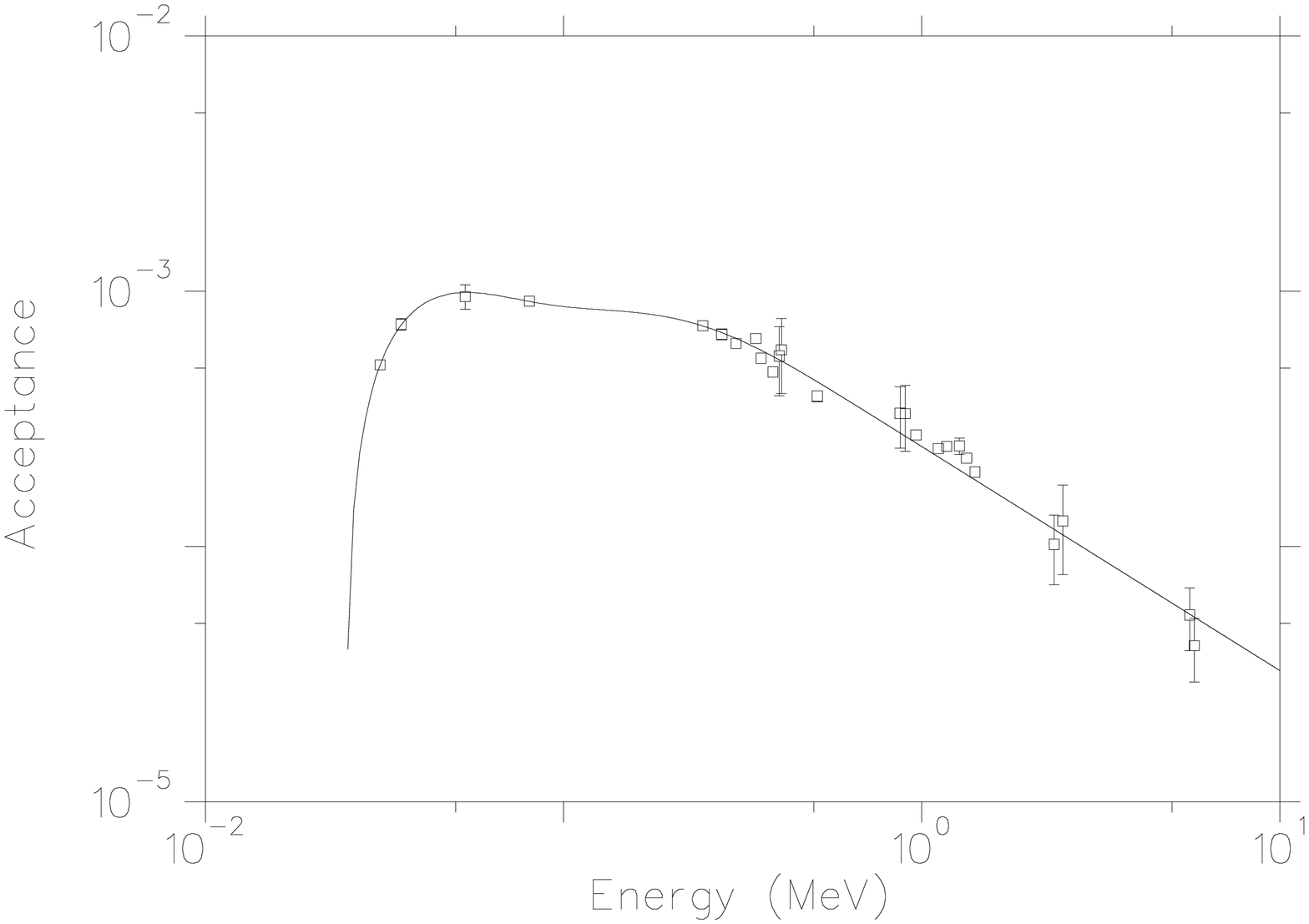,height=12cm,angle=90}
 \caption{The $\gamma$-ray acceptances for HPGe detector 2.}
\label{fig1}
\end{centering}
\end{figure}

The efficiency of the germanium detectors was determined offline with
calibrated sources, and online using muonic X-rays.  An excellent target for
this purpose is gold, and the emission probabilities are given in Table
 ~\ref{tab2}.
The 2p-1s intensities were taken from Hartmann {\it et al.}\cite{r37} and the
others from a Muon Cascade Program \cite{r38}.  The efficiency curve for Ge2 is
presented in Figure~\ref{fig1}.  The individual values were fitted to a curve
of the form: 

\begin{equation}
Efficiency = a_{1}E^{a_{2}} + a_{3}e^{-a_{4}E} + a_{5}e^{-a_{6}E}
+  a_{7}e^{-a_{8}E}   \label{e1}
\end{equation}

where E is the photon energy and a$_{i}$ are parameters for the fit.  This
gives an adequate fit over the whole energy region.  

\section{Data Analysis}
\label{}
Because there are a limited number of $\gamma$-rays which can be produced in
muon capture in $^{14}$N, it was not too difficult to identify the various
lines, using the known transitions.  However, many other transitions were also
observed including: muonic X-rays in C, N, Fe, and Ni; (n,n$^{\prime}$) lines
from N,
Al, Fe, Ge, and Pb; $\beta$-decay lines from $^{16}$N, $^{22}$Na, $^{41}$Ar,
$^{60}$Co, and as mentioned before (n,$\gamma$) capture $\gamma$-rays from H,
N, Al, Cl, Fe, Ge, and In.  Over 150 lines were identified, most with
confidence.  This was done to ensure that we had not missed a candidate line
from muon capture.  The reason that so many background lines occur is twofold.
First in $^{14}$N only 12.7(8)\% of muons capture, the rest decay.  Secondly,
because of the longer muon lifetime,
the $\gamma$-ray gate has to be kept open for $\approx$ 10 $\mu$s, whereas for
an element like Ca a gate of only 1.7 $\mu$s is needed.  This makes the
$^{14}$N spectra about 50 times more vulnerable to background lines than those
for Ca. \\

Another difficulty in $^{14}$N is that the muon capture lines are Doppler 
broadened as they are often emitted while the $^{14}$C recoils after
the neutrino emission.  This broadening is box shaped and amounts to 14 keV 
per MeV $\gamma$-ray energy, i.e. about 100 keV for a 7 MeV $\gamma$-ray. 
A few
energy levels in $^{14}$C and $^{13}$C are long lived with respect to the
slowing
down time ($\approx$ 1 ps), and then the $\gamma$-rays are narrow peaks. In
$^{13}$C the 3684 keV line is Doppler broadened by the neutrino emission and
then again by neutrons of various energies.  Empirically we found a triangular
shape fitted the observed line. \\

The box shaped spectra were fitted to the formula:
   
\begin{equation}
y(x) = \frac{N}{2}\left[erf \left( \frac{E(1+\beta)-x}{\sqrt{2} S} \right)
-erf \left(\frac{E(1-\beta)-x}{\sqrt{2} S} \right) \right] + Ax + B  \label{e2}
\end{equation}

where E is the centroid of the peak, $\beta$ is the recoil velocity divided by
the speed of light ($\approx$ 0.0075), S is the parameter accounting for the
HPGe energy resolution, N is the overall amplitude, and A and B are background
terms.  \\

\begin{figure}
\begin{centering}
\epsfig{figure=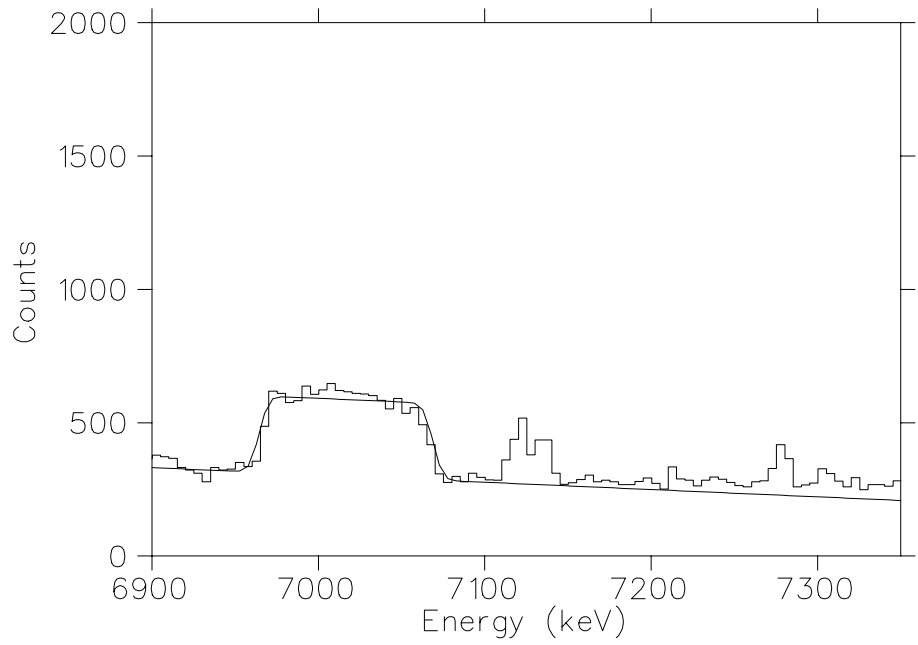,height=12cm,angle=90}
 \caption{The 7017 keV Dopper fit.}
\label{fig2}
\end{centering}
\end{figure}

\begin{figure}
\begin{centering}
\epsfig{figure=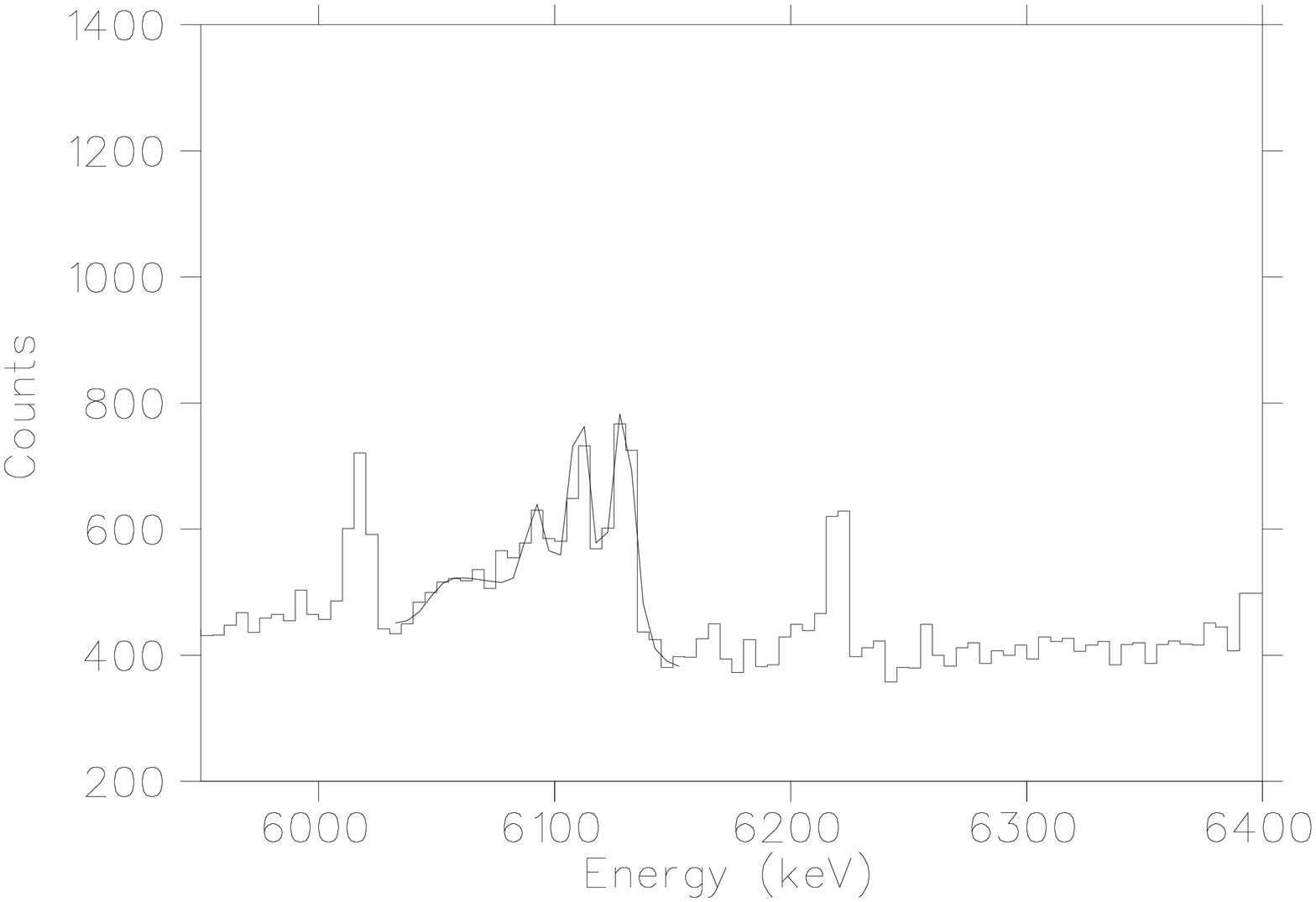,height=12cm,angle=90}
\epsfig{figure=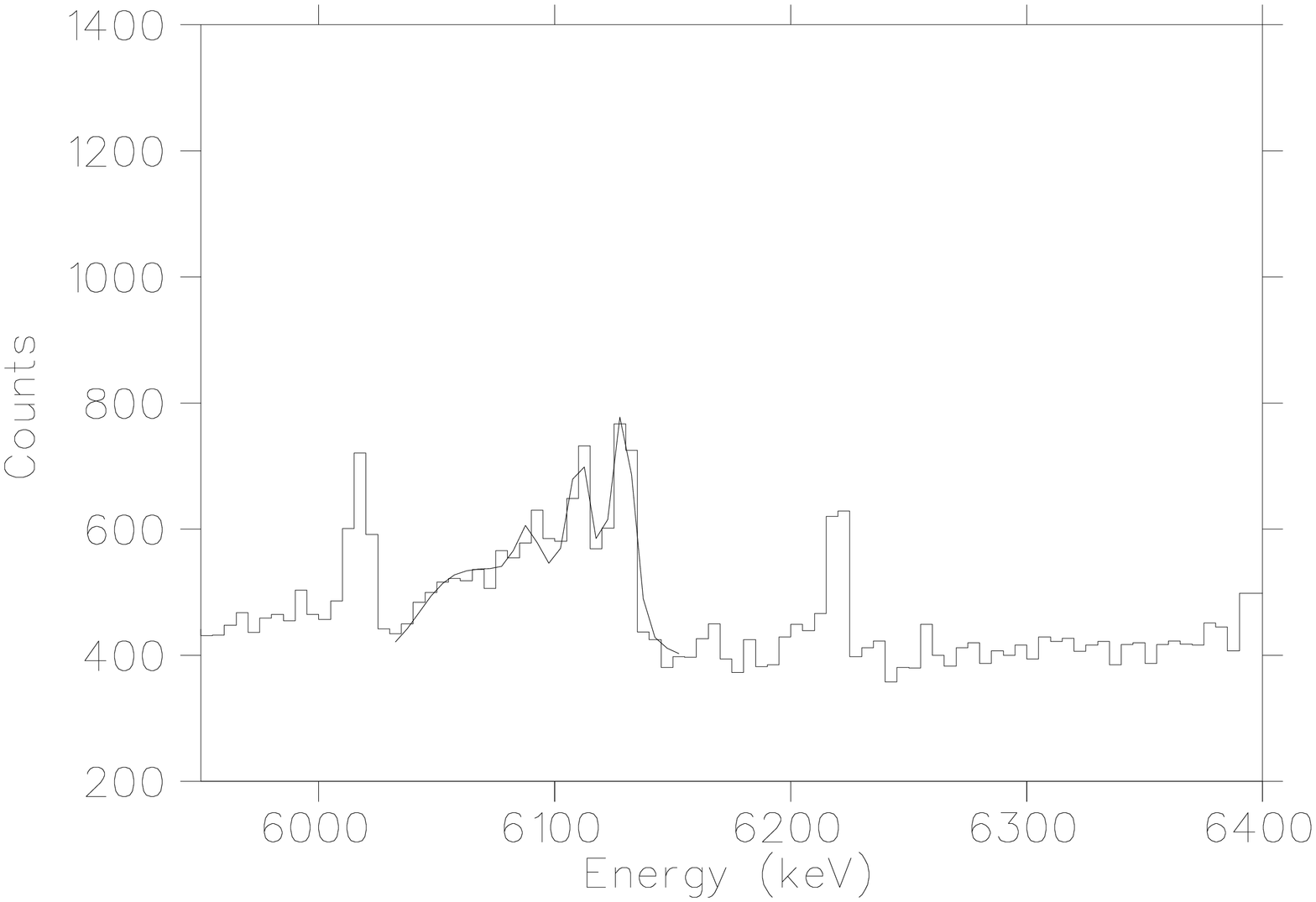,height=12cm,angle=90}
 \caption{Fits to the 6092 keV $\gamma$-ray. (top) with a sloping background
(bottom) with a flat background i.e. A=0 in equation 2.}
\label{fig3}
\end{centering}
\end{figure}

\begin{figure}
\begin{centering}
\epsfig{figure=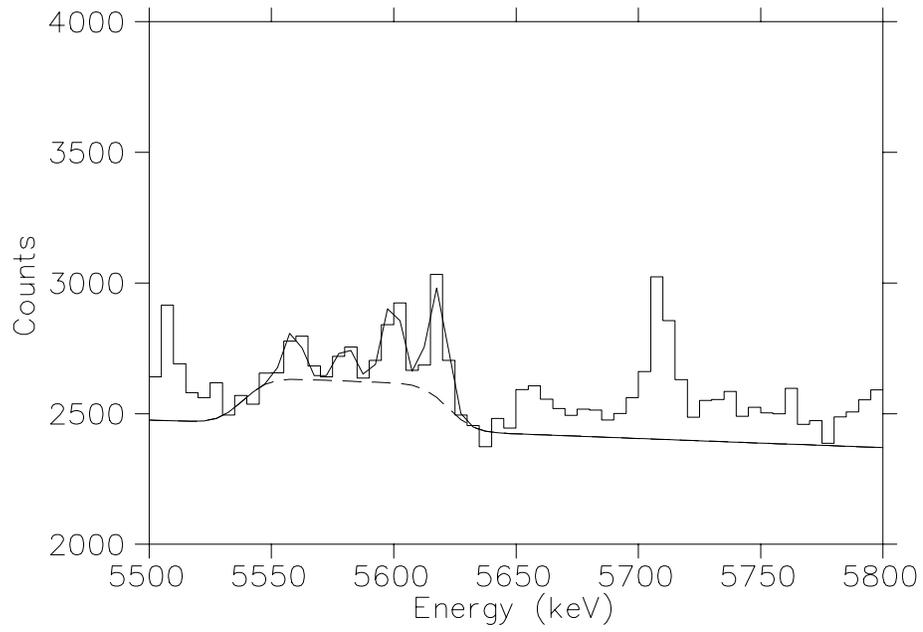,height=12cm,angle=90}
 \caption{The Doppler fit of the 6092 keV, single escape peak.}
\label{fig4}
\end{centering}
\end{figure}

A typical fit to the full energy peak of the 7017 keV peak is shown in
Figure~\ref{fig2}.
This line is fortunately free from background $\gamma$-rays and the yield is
easy to determine (note that there should be a very weak line from
Cl(n,$\gamma$) as predicted by the Cl(n,$\gamma$) branching ratios, the effect
of this line
is subtracted out of the 7017 keV $\gamma$-ray yield).  The 6092 keV peak is
much more complicated and two fits to this peak are illustrated in
Figure~\ref{fig3}.  It is immediately clear
that a spectrum with good statistics is required to fit such a complex
structure. 
We observe two background lines superimposed on the $^{14}$C 6092 keV $\gamma$
-ray, the $^{35}$Cl(n,$\gamma$) at 6110.88 keV and the $^{16}$N $\beta$ decay
line at 6129.14 keV, both fortunately are narrow peaks.  At the centre of the
structure is an interesting  peak, due to $\gamma$-rays from the same $^{14}$C
level at 6092 keV, but narrow because they come from cascades from higher,
longer lived levels.  The single escape peak is illustrated in
Figure~\ref{fig4}, and
another background line is apparent, the 5592.2 keV line from $^{127}$I(n,
$\gamma$) as well as the single escape peaks from all the other effects
of Figure 3.  \\

\begin{figure}
\begin{centering}
\epsfig{figure=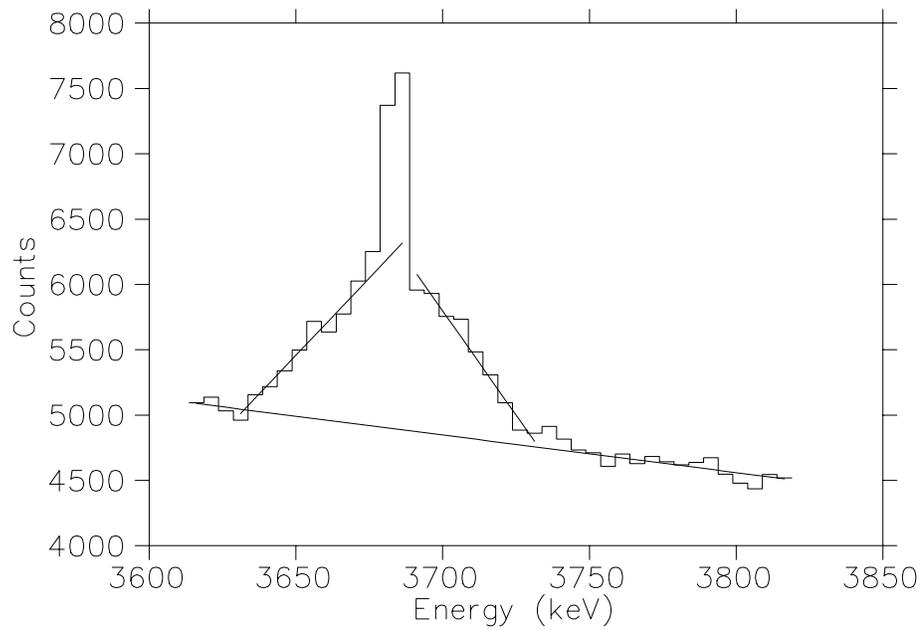,height=12cm,angle=90}
 \caption{The Doppler fit of the 3685 keV peak.}
\label{fig5}
\end{centering}
\end{figure}

In Figure~\ref{fig5} is illustrated a fit to the $^{13}$C peak at 3685 keV.
It is 
composed of a Doppler broadened component, and a central peak, again coming
from a cascade from the long lived level at 3854 keV.  The separation of these
two components is more uncertain, as the shape of the Doppler broadened
component is purely a phenomenological assessment.

\section{Results}
\label{}
From the various fits to the three $^{14}$C $\gamma$-rays (not all illustrated)
we obtain the energies observed in this experiment, listed in Table~\ref{tab3}.
 The
$\gamma$-ray energies are corrected for the recoil correction (1.89, 1.74,
1.42 respectively) to obtain the energy of the excited state.  This is compared
with the compilation of Ajzenberg-Selove \cite{r27}, and a study of the
$^{13}$C(d,p)$^{14}$C reaction by Piskor and Sch\"{a}ferlingova \cite{r39}.  
Although this
study was noted by Ajzenberg-Selove under Reaction 16, it was not integrated 
into her table of level energies which is the same as the 1986 compilation.  
Our results for the 6092 and 6730 keV levels agree, but the 7019 keV is
inconsistent.  We note again that no correction has been made for the escape
peaks being slightly less than the 511 keV difference; our other errors are
much larger. \\

\begin{figure}
\begin{centering}
\epsfig{figure=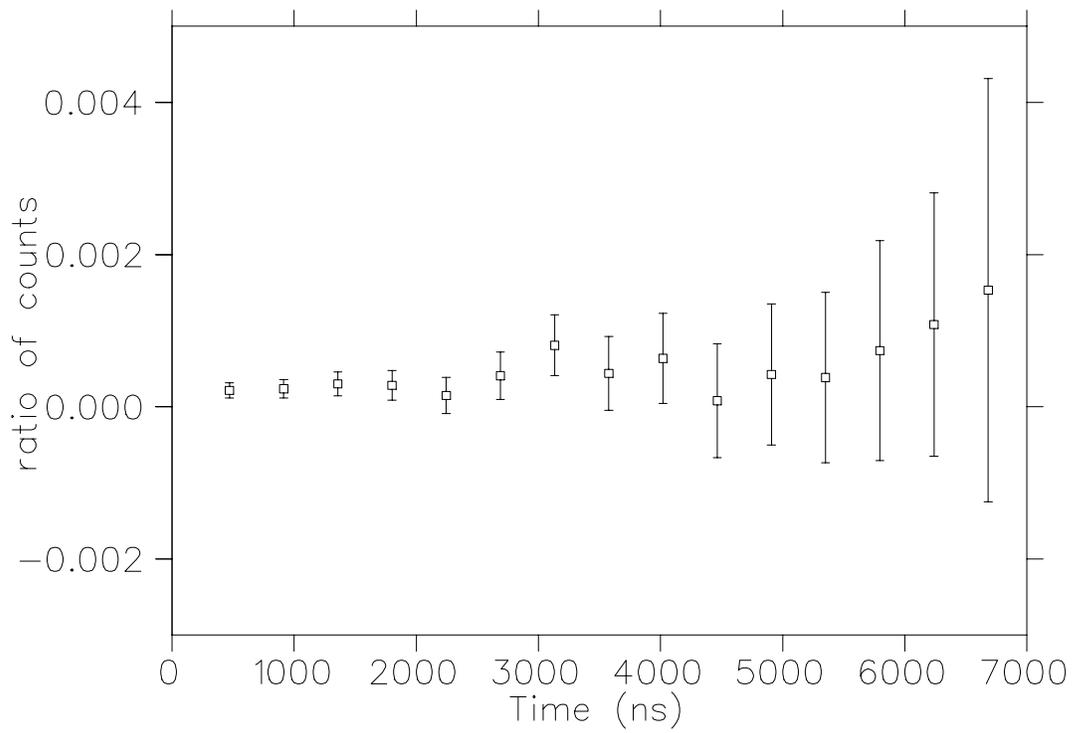,height=14cm,angle=90}
 \caption{The ratio of the 3684 keV $\gamma$-ray to the electron time spectra.}
\label{fig6}
\end{centering}
\end{figure}

As there is concern about the hyperfine transition, we fitted the time
dependence of several $\gamma$-rays.  The 3684 keV line in $^{13}$C has a high
yield, but a complicated shape because of the cascading, see Figure 5.  However
hyperfine asymmetry effects have been seen in such cases, so a study of the
time
dependence was attempted.  The yields of this $\gamma$-ray and of  electron
events
were selected.  Each time spectrum was then fitted to obtain a flat background,
 which was
then subtracted off.  Finally the ratio of these two subtracted spectra is
illustrated in Figure~\ref{fig6}.  Even though about 7000 $\gamma$-ray events
were 
obtained, it is clear that the study is inconclusive.  The results are
compatible with no hyperfine transition with R($\gamma$/e) = 4 $\times$
10$^{-4}$, but could also be fit with a significant 
asymmetry.  Of course there is no guarantee that a particular transition,
especially one from a ($\mu^{-}$,$\nu$n) reaction, will have a large asymmetry.
It is clear that better statistics are needed, and also results from more
lines.
For the rest of the analysis we shall assume no hyperfine transition, and a 
statistical population of the hyperfine levels. \\

The yields of many lines have been studied by using the fitting techniques 
described above, using gaussian, box, or triangular fitting functions, where
appropriate.  In the cases where a limit was set the possible Doppler
broadening of lines was taken into account, where appropiate.  Windows were set
on the expected position, and the nearby background was subtracted.  No
attempt was made to fit possible shapes as these are variable and uncertain.  
However that technique has the potential of setting slightly lower limits.  In
Table ~\ref{tab4} we list the limits on various lines 
obtained using Ge1; also included are the lifetimes of the original levels
\cite{r27,r40,r41}.  Only 3 of these lines have positive identification.  The
4438 keV line in $^{12}$C can be caused by background from target induced
neutrons
in the plastic scintillators.  We made a time study to eliminate background
neutrons, but did
not have enough information to estimate the contribution from target produced
neutrons.  Similarly $^{10}$B could have come from capture on beam
scintillators.  Note that we have no positive identification of
($\mu^{-}$,$\nu$p) nor ($\mu^{-}$,$\nu\alpha$) reactions (i.e. $^{13}$B or
$^{10}$Be) with quite useful limits. \\

In Table ~\ref{tab5} we present the results for the $^{13}$C and $^{14}$C
$\gamma$-rays.
Some of the $^{13}$C results are an average of Ge1 and Ge2, but those for
$^{14}$C are
for Ge1 only as the acceptance for Ge2 was too low to be useful.  The fitting
for the 3685 keV $\gamma$-ray finds that (15.9 $\pm$ 1.0)\% of the
$\gamma$-ray yield is fed from the 3854 keV level via a 169 keV cascade (B.R.
= 36.3\%).  This $\gamma$-ray was below threshold, but from the yield of the
3854 keV ground state transition, using the known branching ratio (62.5\%),
one can deduce that (19 $\pm$ 8 \%) of the 3685 keV yield is from the cascade,
in agreement with the fitting procedure. \\

Equally well the $^{14}$C 6092 keV $\gamma$-ray has a complex structure.  From
the full energy peak we find that (8.6 $\pm$ 5.3) \% of the yield is from
cascade feeding; from the single escape peak the value is (15 $\pm$ 12 \%).
Furthermore there is no evidence for a second Doppler broadened peak on
the principal peak, which we use as evidence against feeding from the 7341 keV
level.  Equally well, the ground state transition for this 7341 keV level
 (B.R. = 16.7 \%)
is not observed.  Thus, for the 6092 keV $\gamma$-ray we assume any
contribution
from the 7341 level to be zero, and average the two central gaussian yields
to obtain that a weighted average of (10 $\pm$ 6)\% of the 6092 keV yield is
cascading.  Thus the direct feeding of the 6094 keV level is (1.2 $\pm$ 0.6)\%
 and the cascading (0.12 $\pm$ 0.08)\% per capture.  Now we know the 6732 keV
level is populated ($\tau$ = 66 ps) and would contribute (0.05 $\pm$ 0.02)\%
per muon capture to the 6092 keV $\gamma$-ray.  This leaves (0.07 $\pm$
0.08)\% which could be feeding from the
6589 keV level which has a branching ratio of 98.9\% for a cascade via 495.35
keV $\gamma$-ray.  This $\gamma$-ray is not observed, but the limit is only 
$<$ 0.049\% per muon capture.  Thus our results are compatible with a feeding
of $\approx$ 0.04\% which is expected in several calculations. \\

From these fitting procedures, and using other known branching ratios we
obtain the direct feeding of all the levels.  Table ~\ref{tab6} presents the
limits
for complex reactions;  Table ~\ref{tab7} presents the results for $^{13}$C
and $^{14}$C.
Apart from the 7019 keV level in $^{14}$C, these are all new results.  The
sum of the yields to bound $^{14}$C levels is (6.0 $\pm$ 1.6)\% which agrees
well with
the ($\pi^{-}$,$\gamma$) result of (6.22 $\pm$ 0.40)\% \cite{r13}, \\

Our absolute yield for the 7019 keV level is lower than all previous results.
Most have not been published and we have to rely on conference proceedings,
reports,
and the review of Mukhopadhyay \cite{r42}.  These results, our own, and a 
recommended average are presented in Table ~\ref{tab8}.  Note that we have
used the rates from other experiments
to calculate the yields; in an experiment it is done the other way but we do
not have the original yields.  As we had significant difficulty with our
absolute normalization, we take the agreement to be satisfactory.  Earlier 
unpublished results could have significant contributions from background lines
and we know that Bellotti {\it et al.} had a serious contamination of unknown origin,
see page 131 of Mukhopadhyay\cite{r41} (we do not observe any line at 6315 keV
).  None of these results had the statistical accuracy, nor the energy
resolution of our own data.  Thus we combine our results with only the Saclay
result to obtain a recommended yield of (6.6 $\pm$ 0.9)\%, i.e. a rate of (4390
$\pm$ 580) s$^{-1}$. \\

All the results for $^{13}$C and $^{14}$C are combined in Table ~\ref{tab9},
and
compared with a variety of reactions, and theoretical calculations.  The
$^{13}$C level feeding is compared with the results for the reaction
$^{14}$N($\gamma$,p) \cite{r46,r47}.  We take the integrated cross section up
to 29 MeV; the ground state and 7550 keV transitions are given by Gellie et
al. \cite{r46}, and the 3089, 3685, and 3854 keV levels are taken from
Thompson {\it et al.}\cite{r47} who studied the $\gamma$-rays emitted during
irradiation by 29 MeV bremsstrahlung.  The (d,$^{3}$He) reaction has been
studied by Hinterberger {\it et al.}\cite{r48} and the spectroscopic factor
C$^{2}$S was found to be 0.8, 0.3, and 1.4 for the ground state, 3685, and
7550 keV states respectively; the feeding of the 3089 keV level is small (
C$^{2}$S $\approx$ 0.05) and the feeding of the 3854 keV level is not even
detected.  Thus the ($\gamma$,p) reaction is a better analogue because the
3089 and 3854 keV levels are clearly detected in muon capture.  Kissener
{\it et al.}\cite{r18} predicted that the pattern of capture be close to the
spectroscopic
factors, with none feeding the 3089 and 3854 keV levels, in contradiction to 
experiment. \\

For $^{14}$C, we compare with the calculation of Kissener
{\it et al.}\cite{r18} for
the ($\mu$,$\nu$) reaction.  Remember that the split between the 7019 keV 
level and the 8318 keV level is phenomenological.  The calculation of
Mukhopadhyay \cite{r9} was for only allowed transitions, i.e. 1$^{+}$
transitions to 0$^{+}$, 1$^{+}$, 2$^{+}$ states.  The 28\% feeding of the 
7019 keV was confirmed by Desgrolard {\it et al.}\cite{r10} for the
Cohen-Kurath
model.  Mukhopadhyay estimated that 
the (2p-2h) contributions would roughly halve the transition rate;
the other half would go to another 2$^{+}$
state and we have used this estimate in Table 9.  Note that Mukhopadhyay gives
transition rates, so we converted to yield using a total capture rate
of (66 $\pm$ 5) $\times$ 10$^{3}$ s$^{-1}$.  However Kissener {\it et al.}
found a
total capture rate of 109 $\times$ 10$^{3}$ s$^{-1}$ and converted to yield 
themselves, so this difference probably explains why the sums for the yields
of Mukhopadhyay are higher than those of Kissener {\it et al.} It is clear,
however,
that even using the Saclay yield of (7.0 $\pm$ 1.0)\%, the Kissener calculation
is still overestimating the yield for the 7012 keV level.  A major problem thus
remains. \\

We also list in Table 9 the experimental results of Perroud
{\it et al.}\cite{r13}
for the ($\pi^{-}$,$\gamma$) reaction at rest.  As their resolution was only
720 keV all the bound levels are in one unresolved peak with a yield of
6.2(4)\% of all radiative capture.  We also give the calculated yields of 
Kissener {\it et al.}\cite{r16} for the ($\pi^{-}$,$\gamma$) reaction.  These
values
have been estimated from the height of the bars in a figure, so are approximate
.  We can see that the calculation fits the experimental yields quite well,
and both compare quite favourably with the ($\mu^{-}$,$\nu$) reaction, even
though only 17\% of the ($\pi^{-}$,$\gamma$) captures are from the 1s state. \\

We can now estimate the overall situation for muon capture in $^{14}$N, using
empirical information.  If we renormalize our results up by 50\% to approach
the Saclay yields, we suggest about 9(2)\% of muon captures give bound states
in $^{14}$C and 14(3)\% produce bound excited states in $^{13}$C.  The
($\gamma$,p) reaction then implies about 32(8)\% direct feeding of the $^{13}$C
ground state and 27(8)\% direct feeding of the $^{12}$C ground state
via the 7.55 level in $^{13}$C.  If we add 4\% feeding of the
first excited state in $^{12}$C and 13\% for charged particles, we obtain a
sum of 99(10)\%, which is reasonable.  Note that this empirical approach is
fairly similar to the theoretical estimate of Kissener {\it et al.} viz. 15\%
to 
bound levels of $^{12}$C, 42\% to bound levels of $^{13}$C and 40\% to bound
levels of $^{12}$C with only 3\% charged particle emission. \\

\section{Conclusions}
\label{}
We have provided a lot of new information about muon capture in $^{14}$N.  The
direct transitions to $^{14}$C are clearly detected with major feeding going
to only three of the 7 bound levels.  These three levels, at 6094, 6732, and
7019 keV have similar yields, yet the calculation of Kissener {\it et al.}
predicts
that the 7019 keV should dominate.  (Note that our relative yields are more
dependable than the absolute values).  It thus seems that the nuclear structure
of the A = 14 system is quite complex and 2p-2h excitations in the wave
function make radical changes to the yields.  Calculations along the lines of
work in $^{12}$C \cite{r5,r6a} and $^{16}$O \cite{r6,r6b} would be really
helpful.
\\

Three $\gamma$-rays from $^{13}$C are observed and their pattern follows the
($\gamma$,p) reaction yields better than the knock out reactions and their
spectroscopic factors.  Even the ($\gamma$,p) reaction is not a perfect 
analogue, and it would be interesting to have data on the integrated yield for
lower energy bremsstrahlung ($\approx$ 24 MeV or so). \\

Other $\gamma$-rays have been searched for but no convincing evidence found.
The
only line clearly detected was from the 4439 keV state in $^{12}$C, and it is
certainly excited in muon capture, but it is also strongly excited by
background mechanisms, so no useful estimate could be made (one would need
extensive runs with carbon and boron targets to separate the various effects).
\\

No $\gamma$-rays were observed from the reaction
$^{14}$N($\mu^{-}$,$\nu$p)$^{13}$B which is surprising.  We obtained a limit
of $<$ 0.17\% for the first excited state at 3483 keV.  In other light nuclei
this reaction has been clearly detected; for example Miller
{\it et al.}\cite{r20}
found that the reaction $^{24}$Mg($\mu^{-}$,$\nu$p)$^{23}$Ne gives a yield of
0.20(5)\% for the 980 keV $\gamma$-ray and 0.5(1)\% for the 1770 keV line;
similarly they observed for $^{28}$Si($\mu^{-}$,$\nu$p)$^{27}$Mg a yield of 
1.9(2)\% for the 984 keV line.  Our only comment is that the high energy of the
first excited state in $^{13}$B may affect the yield adversely.  Equally well
Kissener {\it et al.} estimate a low yield for this reaction, but there is
obviously
some uncertainty in their estimate.  
\\

In heavy nuclei the reaction ($\mu^{-}$,$\nu$pn) is stronger than the
($\mu^{-}$,$\nu$p) reaction \cite{r49}. In 
$^{40}$Ca these reactions have equal $\gamma$-ray yields of about 6\% each
\cite{r50}.  Miller observed a yield of 4.4(6)\% in $^{24}$Mg and 10(1)\% in
$^{28}$Si.  It is thus surprising that our limit for the first excited state
of $^{12}$B at 953 keV is $<$ 0.27\% \\

The mass A=14 system has retained its enigmatic status.  In particular there
is more than sufficient evidence that the Cohen-Kurath model is inadequate and
that the nuclear structure is quite complex.  A comprehensive theoretical study
using modern computational technology would be most welcome. \\

\section{Acknowledgements}
\label{} 
We wish to thank Ulrich Giesen for his help with the Ge detector
acceptances.  We would also like to thank Andrew MacFarlane for his help
with the design of the liquid nitrogen target vessel.  We wish to thank the
Natural Sciences and Engineering Research Council of Canada and similarly the
National Science Foundation in the U.S.A. for support and equipment; we also
thank the staff at TRIUMF and the National Research Council of Canada for 
providing the muon beam facilities.

\clearpage

\begin{table}[t]
\begin{centering}
 \caption{Energies of $\gamma$-rays and muonic X-rays used as calibration for
the HPGe detector }
\begin{tabular}{|ccc|}   \hline
 & Source & Energy (keV) \\ \hline 
Medium Energy & Background from $^{14}$N(n,n$^{\prime}\gamma$) \cite{r26} &
 5104.89
(10) \\ 
 & $^{13}$C line from muon capture \cite{r27} & 3853.194 (19) \\
 & $^{228}$Th \cite{r26} & 2614.533 (13) \\
 & $\mu^{-}$ Au X-ray 3d$_{5/2} \rightarrow$ 2p$_{3/2}$ \cite{r28} &
 2341.21(45) \\
 & $\mu^{-}$ Au X-ray 2p$_{1/2} \rightarrow$ 1s$_{1/2}$ \cite{r28a} &
 5590.7(5)$^*$ \\
 & $\mu^{-}$ I X-ray 2p$_{1/2} \rightarrow$ 1s$_{1/2}$ \cite{r29} &
 3667.361 (35) \\
 & $\mu^{-}$ I X-ray 2p$_{3/2} \rightarrow$ 1s$_{1/2}$ \cite{r29} &
 3723.742(35) \\ \hline
High Energy & $^{56}$Fe(n,$\gamma$) \cite{r30,r31,r32} & 7645.58 (10) \\
 & $^{56}$Fe(n,$\gamma$) \cite{r30,r31,r32} & 7631.18(10) \\
 & Single escape of $^{56}$Fe(n,$\gamma$) & 7134.58(10) \\
 & Single escape of $^{56}$Fe(n,$\gamma$) & 7120.18(10) \\
 & Double escape of  $^{56}$Fe(n,$\gamma$) & 6623.58(10) \\
 & $^{14}$N(n,n$^{\prime}\gamma$) \cite{r26} & 5104.89(10) \\
 & $\mu^{-}$ Au X-ray 2p$_{1/2} \rightarrow$ 1s$_{1/2}$  \cite{r28a} &
 5590.7(5)$^*$ \\  \hline
\end{tabular}
\\
\label{tab1}
${^*}$These energies were taken from Figure 6 of \cite{r28a}.
\end{centering}
\end{table}

\begin{table}[t]
\begin{centering}
 \caption{The muonic x-ray transition probabilities. }
\begin{tabular}{|cccc|}   \hline
Energy (keV) & Element & Transition & Probability \\ \hline 
102.4 & N & 2p $\rightarrow$ 1s & 0.70 $\pm$ 0.07 \\
400.14\cite{r28} & Au & 5g$_{9/2} \rightarrow$ 4f$_{7/2}$ & 0.39 $\pm$ 0.06 \\
405.654\cite{r28} & Au & 5g$_{7/2} \rightarrow$ 4f$_{5/2}$ & 0.30 $\pm$ 0.06 \\
869.98\cite{r28} & Au & 4f$_{7/2} \rightarrow$ 3d$_{5/2}$ & 0.44 $\pm$ 0.04 \\
899.14\cite{r28} & Au & 4f$_{5/2} \rightarrow$ 3d$_{3/2}$ & 0.31 $\pm$ 0.04 \\
2343.44\cite{r28} & Au & 3d$_{5/2} \rightarrow$ 2p$_{3/2}$ & 0.52 $\pm$ 0.08 \\
2474.22\cite{r28} & Au & 3d$_{3/2} \rightarrow$ 2p$_{1/2}$ & 0.29 $\pm$ 0.08 \\
5590.7\cite{r28a}${^*}$ & Au & 2p$_{1/2} \rightarrow$ 1s$_{1/2}$ & 0.333 $\pm$
 0.019 \\
5762.6\cite{r28a}${^*}$ & Au & 2p$_{3/2} \rightarrow$ 1s$_{1/2}$ & 0.559 $\pm$
 0.032 \\ \hline
\end{tabular}
\\
\label{tab2}
${^*}$These energies were taken from Figure 6 of \cite{r28a}.
\end{centering}
\end{table}

\begin{table}[t]
\begin{centering}
 \caption{$^{14}$C $\gamma$-ray full energies (FE), single escape (SE)
energies,  and excited state energies in keV.
Previous work is taken from the compilation by Ajzenberg-Selove \cite{r27} and
the $^{13}$C(d,p)$^{14}$C work of Piskor and Sch\"{a}ferlingova \cite{r39}.}
\begin{tabular}{|cccccc|}   \hline
Peak & Energy & Equivalent & Excitation & Previous  &
 $^{13}$C(d,p)$^{14}$C \\
Type & & Full Energy & Energy & work \cite{r27} & E$_x$ \cite{r39} \\  \hline 
FE & 7016.63(49) & Same & & & \\
SE & 6505.6(15) & 7016.6(15) & &  & \\
Average & & 7016.6(13) & 7018.5(13)  & 7012.0(42)
 & 7011.4(8)  \\ \hline
FE & 6729.68(18) & Same & & & \\
SE & 6220.1(4) & 6731.1(4) & & & \\
Average & & 6729.9(10) & 6731.6(10) & 6728.2(13) &  6731.58(11) \\
\hline
FE model 1 & 6092.4(13) & Same & & & \\
FE model 2 & 6092.2(11) & Same & & & \\
FE model 3 & 6088.2(22) & Same & & & \\
Average & 6091.76(93) & Same & & & \\
SE & 5580.4(17) & 6091.4(17) & & & \\
Average & & 6091.7(14) & 6093.1(14) &  6093.8(2)  & 6094.05(11) \\
\hline
\end{tabular}
\\
\label{tab3}
\end{centering}
\end{table}

\begin{table}[t]
\begin{centering}
 \caption{The $\gamma$-ray yield results for $^{12}$C and $^{10}$B and limits
for $^{10}$Be, $^{12}$B, and $^{13}$B.  Lifetimes of the initial levels are
taken from Ajzenberg-Selove.\cite{r27,r40,r41}}
\begin{tabular}{|cccc|}   \hline
Gamma Energy & Yield/captured $\mu^{-}$ in (\%) & Nuclide & Lifetime
 \cite{r27,r40,r41} \\ \hline
718.3 $^{*}$ & 0.12 $\pm$ 0.04 & $^{10}$B & 1020(5) fs \\
1021.7 & 0.13 $\pm$ 0.07 & $^{10}$B & 7(3) fs \\
4438.03 $^{*}$ & 1 $\pm$ $^{3}_{1}$ & $^{12}$C & 61(4) fs \\ \hline
& & & \\ \hline
219.4 & $<$ 0.047 & $^{10}$Be & 1.1(3) ps \\
2591 & $<$ 0.15 & $^{10}$Be & $<$ 80 fs \\
 & & & 180 $\pm$ 17 fs \\
2811 & $<$ 0.16 & $^{10}$Be &  1.1(3) ps \\
2895 & $<$ 0.086 & $^{10}$Be  & ?  \\ 
3367.4 $^{*}$ & Hidden by $^{12}$C double escape & $^{10}$Be & 180 $\pm$ 17 fs
\\
5957 & $<$ 0.034 & $^{10}$Be &  $<$ 80 fs \\
720.5 & Hidden by $^{10}$B & $^{12}$B & $<$ 50 fs \\
947.11 & $<$ 0.057 & $^{12}$B & $<$ 70 fs \\
953.10 $^{*}$ & $<$ 0.27 & $^{12}$B &  260(40) fs \\
1667.7 & $<$ 0.04  &$^{12}$B & $<$ 70 fs \\
1673.7 & $<$ 0.06  &$^{12}$B & $<$ 50 fs \\
2722.7 & $<$ 0.01 & $^{12}$B &  ? \\
418 & Hidden by $^{127}$I(n,n$^{\prime}\gamma$) & $^{13}$B & 62 (50) fs \\
596 & Hidden by $^{127}$I(n,n$^{\prime}\gamma$) & $^{13}$B & 62 (50) fs \\
3483 $^{*}$ & $<$ 0.17 & $^{13}$B & ?  \\
3713 & $<$ 0.044 & $^{13}$B & $<$ 380 fs  \\
4131 & $<$ 0.62 & $^{13}$B & 62(50) fs  \\  \hline
\end{tabular}
\\
\label{tab4}
$^{*}$ First excited state\\
\end{centering}
\end{table}

\begin{table}[t]
\begin{centering}
 \caption{The $\gamma$-ray yield results for $^{14}$C and $^{13}$C.
The errors in the yield quoted below are only the relative errors and do not
include a 23\% absolute normalization error. The lifetimes
 are taken from Ajzenberg-Selove \cite{r27}.  }
\begin{tabular}{|cccc|}   \hline
& Gamma Energy (keV) & Yield/captured $\mu^{-}$ in (\%) & Lifetime \cite{r27}
\\ \hline
$^{14}$C & & &  \\ \hline
& 495 & $<$ 0.048 & 4.3(6) ps  \\
& 613 & $<$ 0.12 & 160(60) fs  \\
& 634 & $<$ 0.14 & 96(11) ps  \\
& 809 & $<$ 0.50 & 36(4) fs  \\
& 918 & $<$ 0.29 (probably $\approx$ 0.05) & 13(2) fs  \\
& 1248 & Hidden by Ge(n,$\gamma$) & 160(60) fs  \\
& 6092 $^{*}$ & 1.3  $\pm$ 0.3 & $<$ 10 fs  \\
& 6730 & 1.3  $\pm$ 0.2 & 96(11) ps  \\
& 7017 & 3.4  $\pm$ 0.6 & 13(2) fs  \\
& 7339 & $<$ 0.068 & 160(60) fs  \\ \hline
$^{13}$C & & &  \\ \hline
& 169 & Below threshold & 12.4(2) ps \\
& 595 & Hidden by $^{127}$I(n,n$^{\prime}$)& 1.59(13) fs \\
& 764 & $<$ 0.20 & 12.4(2) fs \\
& 3089 & 1.5 $\pm$ 0.3 & 1.55(15) fs \\
& 3685 $^{*}$ & 5.5 $\pm$ 0.2$^{**}$ & 1.59(13) fs \\
& 3854 & 1.9 $\pm$ 0.2$^{**}$ & 12.4(2) ps \\ \hline
\end{tabular}
\\
$^{*}$ This is the entire peak (Doppler plus Gaussian components).\\
$^{**}$ These yields were averaged values from the two detectors.\\
\label{tab5}
\end{centering}
\end{table}

\begin{table}[t]
\begin{centering}
 \caption{The nuclear level yield results for $^{12}$C and $^{10}$B and limits
for $^{10}$Be, $^{12}$B, and $^{13}$B.  }
\begin{tabular}{|ccccc|}   \hline
Level energy (keV) & $\gamma$-ray used & Yield/captured $\mu^{-}$ in (\%) &
 Reaction & Nuclide  \\ \hline
718.35 $^{*}$ & 718.3 & -0.01 $\pm$ 0.25 & ($\mu^-$,$\nu$p3n) & $^{10}$B  \\
1740.15 & 1021.7 & $<$ 0.13 $\pm$ 0.07 & ($\mu^-$,$\nu$p3n) & $^{10}$B  \\
4438.91 $^{*}$  & 4438.03  & 1 $\pm$ $^{3}_{1}$ & ($\mu^-$,$\nu$2n) &$^{12}$C 
 \\ \hline
& & & & \\ \hline
5958.39 & 5956.5 & $<$ 0.034 & ($\mu^-$,$\nu\alpha$) & $^{10}$Be \\
6179.3 & 219.4 & $<$ 0.20 & ($\mu^-$,$\nu\alpha$) & $^{10}$Be \\
 & 2811 & $<$ 0.21 & ($\mu^-$,$\nu\alpha$)  & $^{10}$Be  \\
953.14 $^{*}$ & 953.10  & $<$ 0.27 & ($\mu^-$,$\nu$pn) & $^{12}$B \\
2620.8 & 947.11 & $<$ 0.41 &  ($\mu^-$,$\nu$pn) & $^{12}$B \\
2723 & 2722.7 & $<$ 0.01 & ($\mu^-$,$\nu$pn)  & $^{12}$B \\
3483 $^{*}$ & 3483  & $<$ 0.17 &  ($\mu^-$,$\nu$p) &  $^{13}$B \\ 
3713 & 3713 & $<$ 0.044 & ($\mu^-$,$\nu$p) & $^{13}$B  \\
4131 & 4131 & $<$ 0.91 & ($\mu^-$,$\nu$p) & $^{13}$B   \\  \hline
\end{tabular}
\\
$^{*}$ First excited state\\
\label{tab6}
\end{centering}
\end{table}

\begin{table}[t]
\begin{centering}
 \caption{The nuclear level yield results for $^{14}$C and $^{13}$C.  Note
that the $\gamma$-ray energies in $^{13}$C and $^{14}$C are lower than the
level excitation energies because of the nuclear recoil.}
\begin{tabular}{|cccc|}   \hline
& Level energy (keV) & $\gamma$-ray used & Yield/captured $\mu^{-}$ in (\%)
\\ \hline
$^{14}$C & & & \\
\hline
& 7341 & 613 & $<$ 0.35 \\
& & 7339 & $<$ 0.42 \\
& 7019 & 918 & $<$ 20 \\
&  & 7017 & 3.4  $\pm$ 1.4 \\
& 6903 & 809 & $<$ 0.50 \\
& 6589 & 495 & $<$ 0.049 \\
& 6732 & 634 & $<$ 3.9 \\
&  & 6730 & 1.40  $\pm$ 0.49 $\pm$ $^{0}_{0.12}$  \\ 
& 6094 & 6092 & 1.2  $\pm$ 0.6 $\pm$ $^{0}_{0.79}$ \\ \hline
$^{13}$C & & & \\ \hline
& 3853 & 3853 & 3.0 $\pm$ 0.8 \\ 
& 3684 & 3684 & 4.4 $\pm$ 1.3 \\
& 3089 & 3089 & 1.4 $\pm$ 0.7 \\  \hline
\end{tabular}
\\
\label{tab7}
\end{centering}
\end{table}

\begin{table}[t]
\begin{centering}
 \caption{Measurements of partial muon capture rates in $^{14}$N to the 7019
keV level.  Note the uncertainty on the total capture rate of (66 $\pm$ 5)
$\times$ 10$^{3}$ s$^{-1}$ contributes significantly to errors. }
\begin{tabular}{|ccc|}   \hline
Experiment & Partial capture rate (s$^{-1}$) & Yield/captured $\mu^{-}$ in (\%)
\\ \hline
Present & 2240 $\pm$ 920 & 3.4 $\pm$ 1.4 \\
Saclay \cite{r7} & 4640 $\pm$ 700 & 7.0 $\pm$ 1.0 \\
Dubna \cite{r43} & 10000 $\pm$ 3000 & 15 $\pm$ 5 \\
Louvain \cite{r47} & 8000 & 12\\
PSI (SIN) \cite{r45} & 6000 $\pm$ 1500 & 9.1 $\pm$ 2.3 \\
Recommended value & 4390 $\pm$ 580 & 6.6 $\pm$ 0.9 \\ \hline
\end{tabular}
\\
\label{tab8}
\end{centering}
\end{table}

\begin{table}[t]
\begin{centering}
 \caption{Comparison of $\gamma$-ray yields from muon capture in $^{14}$N with
other reactions.  The direct feeding for each excited state is given, taking
into account the $\gamma$-ray branching ratios and cascade feeding. (All bound
states of $^{13}$C and $^{14}$C are listed plus a few major unbound ones).
Also given are the integrated yields for the ($\gamma$,p) reaction up to 
29 MeV \cite{r46,r47}, the ($\pi^{-}$,$\gamma$) yields of Perroud
{\it et al.}\cite{r13}, the ($\mu^{-}$,$\nu$) calculation of Kissener
{\it et al.}\cite{r18}
 (all transitions), the calculation of Mukhopadhyay \cite{r9}
(allowed transitions only, i.e. 1$^{+}$), and the ($\pi^{-}$,$\gamma$) 
calculation of Kissener {\it et al.}\cite{r16} (all transitions). [Note unbound
energies do not always agree!] }
\begin{tabular}{|ccccccc|}   \hline
Nuclide & Excited State & ($\mu$,$\nu$) direct & ($\mu$,$\nu$) calc & 
($\mu$,$\nu$) calc & ($\gamma$,p) & \\
& (keV) J$^{\pi}$ & feeding (\%) & (\%) \cite{r18} & (\%) \cite{r9} &
 $\int\sigma$.dE &  \\
& & & & & (MeV$\cdot$mb) &  \\  \hline
$^{13}$C & g.s. $\frac{1}{2}^{-}$ & ? & 26 & & 20 & \\ 
 & 3089 $\frac{1}{2}^{+}$ & 1.4(7) & - & & small &  \\ 
 & 3685 $\frac{3}{2}^{-}$ & 4.4(13) & 16 & & 7(2) &  \\ 
 & 3854 $\frac{5}{2}^{+}$ & 3.0(8) & - & & 1.7(7)  &\\ 
 & 7550 $\frac{5}{2}^{-}$ & unbound & 18 & & 17 & \\ \hline
 & & & & & ($\pi^{-}$,$\gamma$) & ($\pi^{-}$,$\gamma$) \\
 & & & & & exp \cite{r13} & calc. \cite{r16} \\
 & & & & & (\%) & (\%)  \\ \hline
$^{14}$C & g.s. 0$^{+}$ & ? & & 0.4 & 0.25(11) & 0.25 \\ 

&
$
\begin{array}{c}
\mbox{6094 1$^{-}$} \\
\mbox{6589 0$^{+}$} \\ 
\mbox{6732 3$^{-}$} \\
\mbox{6903 0$^{-}$} \\ 
\mbox{7019 2$^{+}$} \\
\end{array}
$
&
$
\begin{array}{c}
\mbox{1.2 $\pm$ $^{0.6}_{1.0}$ } \\
\mbox{$<$ 0.05} \\ 
\mbox{1.4(5)} \\
\mbox{$<$ 0.5} \\ 
\mbox{3.4(14)} \\
\end{array}
$
&
$
\begin{array}{c}
\mbox{1.1} \\
  \\ 
\mbox{2.7 } \\
  \\ 
\mbox{11 } \\
\end{array}
$
&
$
\begin{array}{c}
  \\
\mbox{0.3} \\ 
   \\
   \\ 
\mbox{15} \\
\end{array}
$
&
$
\left\}
\begin{array}{c}
  \\
  \\ 
\mbox{= 6.2(4)} \\ 
   \\ 
   \\
\end{array}
\right.\
$
&
$
\begin{array}{c}
  \\
  \\ 
\mbox{3.4} \\
  \\ 
  \\
\end{array}
$
\\

  & 7341 2$^{-}$ & $<$ 0.4 & & & & 0.8 \\ 
  & 8318 (2$^{+}$) & unbound & 11 & 15 & 3.4(3) & 3.1 \\ 
  & 11306 (1$^{+}$) & unbound &  & 2 & 2.3(4) & 1.3 \\ 
 \hline
\end{tabular}
\\
\label{tab9}
\end{centering}
\end{table}


\begin{thebibliography}{60}

\bibitem{r1} T.P. Gorringe {\it et al.}, Phys. Rev. $\underline{C60}$ (1999)
055501. 
\bibitem{r2} B. Johnson {\it et al.}, Phys. Rev. $\underline{C54}$ (1996) 2714.
\bibitem{r3} B.A. Brown, A. Etchegoyen, W.D.M. Rae, and N.S. Godwin.  The
Oxford-Buenos-Aires-MSU shell-model code (OXBASH).  MSUCL Report N$^{O}$ 524,
(1986).
\bibitem{r4} B.A. Brown and B.H. Wildenthal, Ann. Rev. Nucl. Part. Sci.
$\underline{38}$ (1998) 29.
\bibitem{r5} A.C. Hayes and I.S. Towner, Phys. Rev. $\underline{C61}$ (2000)
044603.
\bibitem{r6a} C. Volpe {\it et al.}, Phys. Rev. $ \underline{C62}$ (2000)
 015501.
\bibitem{r6} W.C. Haxton and C. Johnson, Phys. Rev. Lett. $\underline{65}$
(1990) 1325.
\bibitem{r6b} E.K. Warburton, I.S. Towner, and B.A. Brown, Phys. Rev.
 $\underline{C49}$ (1994) 824.
\bibitem{r7} M. Giffon {\it et al.}, Phys. Rev. $\underline{C24}$ (1981) 241.
\bibitem{r8} T. Suzuki {\it et al.}, Phys. Rev. $\underline{C35}$ (1987) 2212.
\bibitem{r9} N.C. Mukhopadhyay, Phys. Lett. $\underline{44B}$ (1973) 33.
\bibitem{r10} P. Desgrolard {\it et al.}, Nuovo Cim. $\underline{A43}$ (1979)
 120.
\bibitem{r11} P. Desgrolard and P.A.N. Guichon, Phys. Rev. $\underline{C19}$
(1978) 475.
\bibitem{r12} V. Wiaux, Ph.D. Thesis, Universit\'{e} catholique de Louvain,
1999.
\bibitem{r13} J.P. Perroud {\it et al.}, Nucl. Phys. $\underline{A453}$ (1986)
 542.
\bibitem{r14} M. Gmitro {\it et al.}, Sov. J. Particles Nuclei $\underline{13}$
(1982) 513; ibid $\underline{14}$ (1983) 323.
\bibitem{r15} H.R. Kissener {\it et al.}, Nucl. Phys. $\underline{A302}$
 (1978) 523.
\bibitem{r16} H.R. Kissener {\it et al.}, Nucl. Phys. $\underline{A312}$
 (1978) 394.
\bibitem{r17} G.A. Needham {\it et al.}, Nucl. Phys. $\underline{A385}$
 (1982) 349.
\bibitem{r17a} K.P. Jackson, private communication.
\bibitem{r18} H.R. Kissener {\it et al.}, Nucl. Phys. $\underline{A215}$
 (1973) 424.
\bibitem{r19} D.F. Measday, Phys. Reports, to be published.
\bibitem{r20} G.H. Miller {\it et al.}, Phys. Rev. $\underline{C6}$ (1972) 487.
\bibitem{r21} G. Culligan, J.F. Lathrop, V.L. Telegdi, R. Winston and R.A.
Lundy, Phys. Rev. Lett. $\underline{7}$ (1961) 458.
\bibitem{r22} R. Winston, Phys. Rev. $\underline{129}$ (1963) 2766.
\bibitem{r23} T.J. Stocki, Ph.D. thesis, University of British Columbia 1998;
to be published.
\bibitem{r24} K. Ishida {\it et al.}, Phys. Lett. $\underline{167B}$ (1986) 31.
\bibitem{r26} R.B. Firestone {\it et al.}, Table of Isotopes, 8$^{th}$ Edition,
published by J. Wiley, N.Y., March 1996.
\bibitem{r27} F. Ajzenberg-Selove, Nucl. Phys. $\underline{A523}$ (1991) 1.
\bibitem{r28} R. Engfer {\it et al.}, Atomic Data and Nuclear Data Tables
$\underline{14}$ (1974) 509. 
\bibitem{r28a} R.J. Powers {\it et al.}, Nucl. Phys. $\underline{A230}$ (1974)
 413. 
\bibitem{r29} G. Fricke {\it et al.}, Atomic Data and Nuclear Data Tables 
$\underline{60}$ (1995) 177.
\bibitem{r30} World Wide Web site: Thermal Neutron Capture Gammas by Energy:
http://www.nndc.bnl.gov/wallet/tncngtblcontentbye.shtml
\bibitem{r31} M.R. Bhat, Nuclear Data Sheets $\underline{67}$ (1992) 195 (see
p230).
\bibitem{r32} R. Vennink {\it et al.}, Nucl. Phys. $\underline{A344}$ (1980)
 421.
\bibitem{r33} A. H. Wapstra, Nucl. Inst. Meth. $\underline{A292}$ (1990) 671.
\bibitem{r34} World Wide Web site: http://www.nndc.bnl.gov
\bibitem{r35} D.R. Tilley {\it et al.}, Nucl. Phys. $\underline{A564}$ (1993)
 1.
\bibitem{r36} E.K. Warburton and D.E. Alburger, Nucl. Phys. $\underline{A385}$
(1982) 189.
\bibitem{r37} F.J. Hartmann {\it et al.}, Z. Phys. $\underline{A305}$ (1982)
 189.
\bibitem{r38} V.R. Akylas and P. Vogel, Computer Physics Communications 
$\underline{15}$ (1978) 291.
\bibitem{r39} S. Piskor and W. Sch\"{a}ferlingova, Nucl. Phys.
$\underline{A510}$ (1990) 301.
\bibitem{r40} F. Ajzenberg-Selove, Nucl. Phys. $\underline{A490}$ (1988) 1.
\bibitem{r41} F. Ajzenberg-Selove, Nucl. Phys. $\underline{A506}$ (1990) 1.
\bibitem{r42} N.C. Mukhopadhyay, Physics Reports $\underline{30C}$ (1977) 1.
\bibitem{r43} A.I. Babaev {\it et al.}, JINR Report R-14-42-41 (1968) quoted
 by V.A. Vartanyan {\it et al.}, Sov. J. Nucl. Phys. $\underline{11}$ (1970)
 295.
\bibitem{r44} L. Grenacs, quoted on page 57 of Mukhopadhyay \cite{r42}. 
\bibitem{r45} E. Bellotti {\it et al.}, SIN Newsletter $\underline{1}$ (1976)
 41 and quoted on pages 58 and 131 of Mukhopadhyay \cite{r42}.
\bibitem{r46} R.W. Gellie {\it et al.}, Can. J. Phys. $\underline{50}$ (1972)
 1689.
\bibitem{r47} M.N. Thompson {\it et al.}, Phys. Lett. $\underline{B31}$ (1970)
 211.
\bibitem{r48} F. Hinterberger {\it et al.}, Nucl. Phys. $\underline{A106}$
 (1968) 161.
\bibitem{r49} A. Wyttenbach {\it et al.}, Nucl. Phys. $\underline{A294}$
 (1978) 278.
\bibitem{r50} P. Igo-Kemenes {\it et al.}, Phys. Lett. $\underline{B34}$
 (1971) 286.

\end{thebibliography}
\end{document}